\documentstyle[11pt,epsfig]{article}
\renewcommand{\title}[1]{{\Large\bf\mbox{}\\\medskip#1\bigskip\medskip\\}}
\renewcommand{\author}[1]{{\large #1\smallskip\\}}
\newcommand{\address}[1]{{\em #1\medskip\\}}
\def\be{\begin{equation}}
\def\ee{\end{equation}}
\def\bea{\begin{eqnarray}}
\def\eea{\end{eqnarray}}
\def\ba{\begin{array}}
\def\ea{\end{array}}

\def\0{$\Gamma_0$}

\def\p{\phi}

\def\t{\theta}

\begin{document}
\begin{center}

\title{Generalized Fibonacci numbers  and dimer statistics}
\author{W. T. Lu and F. Y. Wu}
\address{Department of Physics,
Northeastern University, Boston, Massachusetts 02115}

\begin{abstract}
We establish new product identities involving the $q$-analogue of the Fibonacci numbers.
   We show that the identities  lead to alternate expressions
of generating functions for close-packed dimers on non-orientable 
surfaces. 
\end{abstract}

\end{center}


The recursion relation
\be
{\cal F}_{n+2}={\cal F}_{n}+{\cal F}_{n+1} \label{fib}
\ee
with the initial values ${\cal F}_0 ={\cal F}_1=1$ produces the
well-known Fibonacci (1170-1250) numbers
$ 1,\ 1,\ 2,\ 3,\ 5,\ 8,\ 13,\ 21,\ 34,\ 55,\ \cdots$\ . 
The Fibonacci sequence can be generalized in a number of ways 
\cite{andrews,gL}.
One generalization is the $q$-analogue of the sequence defined by
 the  recursion relation
\be
{\cal F}_{n+2}(q)={\cal F}_{n}(q)+q{\cal F}_{n+1}(q)\ ,\label{qfib}
\ee
and the initial condition ${\cal F}_0(q)=1$ and ${\cal F}_1(q)=q$.
Here $q$ is any real or complex number.  This yields  the sequence
$1,\ q,\ q^2+1,\ q^3+2q,\ q^4+3q^2+1,\ \cdots $\ .
 
It can be readily established that the generating 
function of  ${\cal F}_n(q)$ is   
\be
{1\over 1-qs-s^2}=\sum_{n=0}^\infty {\cal F}_n(q)s^n \ .\label{genf}
\ee
Writing $
q=x-x^{-1}$ 
and partial fractioning 
the left-hand side of (\ref{genf}), one obtains
the explicit expression
 \be
{\cal F}_n(q)={x^{n+1}+(-1)^nx^{-(n+1)}\over x+x^{-1}}\ 
, \quad\quad n=0,1,2,\cdots . \label{expand}
\ee


 We have  following identities which we state as a theorem. 

\noindent
{\it Theorem:
(A)  For $M=$ integers and $\t_m ={m\pi/( 2M+1)}$, we have
\be
{\cal F}_{M}(q)\pm i\ {\cal F}_{M-1}(q) 
=\prod_{m=1}^{M}\Big(q\ \mp 2\ i\ (-1)^{m}\cos\t_{m}\Big),\quad
M=1,2,\cdots .\label{prop1} 
\ee
 
(B) For   $M=$ integers we have
\be
{\cal F}_M(q)=\prod_{m=1}^M\Big(q+2i\cos {m\pi\over M+1}\Big),\quad
M=1,2,\cdots .\label{propB} 
\ee

(C) For   $N=$ integers and  $\phi_n  ={(2n-1)\pi}/{2N}$,
 we have 
\bea
{\cal F}_{N}(q)+{\cal F}_{N-2}(q)
&=&\prod_{n=1}^{N}\Big(q\pm 2i\cos \p_n\Big) \hskip 2.7cm N= 2,4,\cdots \label{prop2a} \\
&=&\pm 2i +\prod_{n=1}^{N}\Big(q\pm 2i(-1)^{n}\sin\p_n\Big),\
 \quad N=1,3,\cdots . \label{prop2b}
\eea

(D)  For $M=$ integers, $N$ = odd, $ q_n=2\tau^{-1}\sin \phi_n$, and $ p_m = 2\tau\cos \theta_m $,
 we have
\be
\tau^{MN}\prod_{n=1}^{N}\Big({\cal F}_{M}(q_n)
-i\ (-1)^{n+M}{\cal F}_{M-1}(q_n)\Big)
=i^{-M^2}\prod_{m=1}^{M}
\Big({\cal F}_{N}(p_m)+{\cal F}_{N-2}(p_m)
+2i(-1)^{m+M}\Big),\label{id}\\
\ee
 } 

These identities can be established by determining the zeroes of the expressions.
To establish $(A)$,  for example, using  (\ref{expand}) it is straightforward to deduce the identity 
 \be
{\cal F}_{M}(q)\pm\ i\ {\cal F}_{M-1}(q)
 = { x^{-M}}\Bigg( {x^{2M+1}-  (\pm i)^{2M+1}\over x\mp i}
\Bigg) \ .
\label{FF-left}
\ee
The quantity inside the parentheses in (\ref{FF-left})
is a polynomial of degree $2M$ in $x$ with the
highest term $x^{2M}$.  It has $2M$ zeroes at
 \bea
x_m=\pm\ i\ e^{i \t_{2m}}, \hskip 1.5cm m=1,2,\cdots, 2M . \nonumber
\eea
 Using the fact that $x_{2M} = -x_1^*,\ 
x_{2M-1} = -x_2^*,\ ..., x_{M+1} = -x_M^*$, we obtain 
 \bea
{\cal F}_{M}(q)\pm\ i\ {\cal F}_{M-1}(q)&=&
x^{-M}\prod_{m=1}^{2M}(x-x_m)\nonumber \\
&=& x^{-M} \prod_{m=1}^{M}(x - x_m)(x + x_m^*) \nonumber \\
&=& \prod_{m=1}^{M} \Big(q\mp 2\ i\ \cos \t_{2m}\Big) \ . \label{p1}
\eea
This leads to $(A)$ after we make use of the identities
$\cos \t_{2M} = -\cos \t_1, \ \cos \t_{2M-2} = (-1)^3 \cos \t_3,\cdots$.
 The identity $(B)$ is proved in a similar fashion.

To establish $(C)$ we note that for all $N$ we have 
\bea
{\cal F}_{N}(q)+{\cal F}_{N-2}(q) =x^{N}+(-x)^{-N} \nonumber.
\eea
For $N=$ even, the  polynomial  $x^{-N}(x^{2N}+1)$  has $2N$ zeroes at
\be
x_n=\mp ie^{i\p_n},\quad x_n=\mp ie^{-i\p_n},\quad n=1,2, \cdots, N.
\ee
This leads to (\ref{prop2a}) by factorizing the polynomial as in the proof of $(A)$.

For $N=$ odd, we have instead
 \bea
{\cal F}_{N}(q)+{\cal F}_{N-2}(q) \mp 2i &=&x^{N}-x^{-N}\mp 2i  \nonumber  \\
&=& x^{-N} \Big( x^N \mp i \Big)^2  \nonumber \\
&=&  x^{-N}\prod_{n=1}^N (x-x_n)^2 
\eea
where $x_n=\mp (-1)^{n}e^{i\p_n}.$ 
  The identity (\ref{prop2b}) now follows 
the replacement of
  $x_1=-x_N^*, \ x_2=-x_{N-1}^*, ...$ in one of the
$x-x_n$ factors.  
 
To establish $(D)$, write
$q=q_n=2\tau^{-1}\sin  \p_n$ in $(A)$ and  $q=p_m
= 2\tau \cos\t_m $ in $(C)$ to obtain, respectively,  
\bea
{\cal F}_{M}(q_n)- i(-1)^{n+M} {\cal F}_{M-1}(q_n) &=&
\prod_{m=1}^{M}\Big( 2 \tau^{-1} \sin\phi_n 
\ + 2\ i\ (-1)^{m+n+M}\cos\t_{m}\Big) \nonumber \\ 
 {\cal F}_{N}(p_m)+{\cal F}_{N-2}(p_m)
+ 2i(-1)^{m+M}&=&\prod_{n=1}^{N}\Big(2\tau \cos\t_m
-2i(-1)^{n+m+M}\sin\p_n\Big). \nonumber
 \eea
The identity $(D)$  now follows by combining these two expressions and the fact
that,  for $N=$ odd integers, 
 \bea
\prod_{m=1}^{M}\prod_{n=1}^{N}\Big(i(-1)^{m+M+n+1}\Big)=i^{M^2}.
\label{id-i} \nonumber
\eea

The identities $(A)$ - $(D)$ can be used to deduce alternate expressions 
for  dimer generating functions on a  M\"obius strip.  Let 
$Z^{\rm Mob}_{\cal M,N}(z_h,z_v)$ be the
generating function on  an ${\cal M} \times {\cal N}$  M\"obius strip of width ${\cal N}$
with horizontal dimer weights $z_h$ and vertical dimer weights $z_v$.
It can be shown that, by using  $(A)$ - $(D)$ 
and the equivalences of the sets 
$\{\sin\p_{2n}\}=\{\cos\p_n\}$ for $N=$ even and
$\{\sin\p_{2n}\}=\{(-1)^n\sin\p_n\}$ with $n=1,2,\cdots,N$,
one establishes the following equivalent expressions:

For $N=$ even  and  $\tau=z_v/z_h$,
\bea
Z^{\rm Mob}_{2M,{N}}(z_h,z_v)
&=& \prod_{m=1}^{M}\prod_{n=1}^{N/2}\Big(
   4z_h^2\sin^2\p_{2n}+4z_v^2\cos^2\t_m\Big) \label{ee1} \\
&=& z_h^{MN}\prod_{m=1}^{M}\Big({\cal F}_{N}(p_m)
+{\cal F}_{N-2}(p_m)\Big) \label{ee2} \\
&=&  z_v^{MN}\prod_{n=1}^{N/2}{\cal F}_{2M}(q_n), \hskip 2cm N = {\rm even} \label{ee3} \\
Z^{\rm Mob}_{2M-1,{N}}(z_h,z_v)
&=&2 z_h^{N/2}\prod_{m=1}^{M-1}\prod_{n=1}^{N/2}\Big(
  4z_h^2\sin^2\p_{2n}+4z_v^2\cos^2(m\pi/ 2M)\Big) \label{oe1}\\
&=&2z_h^{(2M-1)N/2}\prod_{m=1}^{M-1}\Big[{\cal F}_{N}\Big(2\tau \cos(m\pi/2M)\Big)
+{\cal F}_{N-2}\Big( (2\tau \cos(m\pi/2M)\Big)\Big]   \label{oe2} \\
&=&{2}z_h^{N/2}z_v^{(M-1)N}\prod_{n=1}^{N/2}\Big(q_n^{-1}{\cal F}_{2M-1}(q_n)\Big),
\hskip 1cm N={\rm even} \label{oe3}.
\eea
For $N=$ odd and  $\tau=z_v/z_h$, 
\bea
Z^{\rm Mob}_{2M,{N}} (z_h,z_v)
&=& {\rm Re}\Bigg[(1-i)\prod_{m=1}^{M}\prod_{n=1}^{N}\Big(
   2i(-1)^{M+m+1}z_h\sin\p_{2n}+2z_v\cos\t_m\Big)\Bigg] \label{eo1} \\
 &=& z_h^{MN}\, 
{\rm Re}\Bigg[(1-i)\prod_{m=1}^{M}\Big(
{\cal F}_{N}(p_m)+{\cal F}_{N-2}(p_m)
+2i(-1)^{m+M}\Big)\Bigg] \label{eo2} \\
  &=&z_v^{MN}\, {\rm Re}\Bigg[(1-i)i^{M^2}\prod_{n=1}^{N}\Big(
   {\cal F}_M(q_n)-i(-1)^{n+M}{\cal F}_{M-1}(q_n)\Big)\Bigg], 
\quad N={\rm odd}  .  \label{eo3}
\eea
 Expressions (\ref{ee1}), (\ref{oe1}) and (\ref{eo1}) are those given by Lu and Wu \cite{luwu99,luwu02},
expressions (\ref{ee2}), (\ref{oe2})  and (\ref{eo2}) are given by Tesler \cite{tesler}, and
expressions (\ref{ee3}), (\ref{oe3})  and (\ref{eo3}) are new.

Work has been supported in part by NSF grant DMR-9980440.

\end{document}